\documentclass[twocolumn]{aa}

\usepackage{graphicx}
\usepackage{natbib}
\bibpunct{(}{)}{;}{a}{}{,}

\newcommand{\gx}{GX\,13+1}

\begin{document}

\title{Simultaneous radio and X-ray observations \\ of the low-mass X-ray binary \gx}

\subtitle{}

\author{Jeroen Homan \inst{1} \and Rudy Wijnands \inst{2} \and Michael P.\  Rupen \inst{3} \and Rob Fender \inst{4} \and Robert M. Hjellming \inst{5} \and Tiziana di Salvo \inst{4} \and Michiel van der Klis \inst{4} }

\institute{INAF - Osservatorio Astronomico di Brera, Via E. Bianchi 46, I-23807 Merate, Italy
\and
School of Physics and Astronomy, University of St Andrews, St Andrews, Fife KY16 9SS, Scotland, UK
\and
National Radio Astronomy Observatory, Socorro, NM 87801
\and
Astronomical Institute 'Anton Pannekoek', University of Amsterdam, Kruislaan 403, 1098 SJ, Amsterdam, The Netherlands
\and
Deceased
}

\offprints{Jeroen Homan, \email{homan@merate.mi.astro.it}}

\date{Received / Accepted}

\abstract{We present the results of two simultaneous X-ray/radio
observations of the low-mass X-ray binary GX 13+1, performed in July/
August 1999 with the Rossi X-ray Timing Explorer and the Very Large
Array. In X-rays the source was observed in two distinct spectral  states; a soft state, which had a corresponding 6 cm flux
density of $\sim$0.25 mJy, and a hard state, which was much brighter
at 1.3--7.2 mJy. For the radio bright observation we measured a delay
between changes in the X-ray spectral hardness and the radio
brightness of $\sim$40 minutes, similar to what has been found in the
micro-quasar GRS 1915+105. We compare our results with those of GRS 1915+105 and the atoll/Z-type neutron star X-ray binaries. Although it has some properties that do not match with either atoll or Z sources, GX 13+1 seems more similar to the Z sources. 

 }

\maketitle

\section{Introduction}\label{sec_intro}

Based on their correlated spectral and variability properties, the
brightest persistent neutron star low-mass X-ray binaries (LMXBs) are
often divided in two groups: the atoll and Z sources
\citep{hava1989,va1995a}, after the tracks they trace out in X-ray
colour-colour diagrams (CDs). Although atoll and Z sources share some
variability and spectral properties (in the X-ray band), there are
significant differences between the two groups: atoll sources are
less luminous, have harder X-ray spectra and show stronger rapid time
variability than the Z sources \citep{va1995a}. Atoll sources are
also less luminous in the radio \citep{fehe2000}. Although some of
the observational differences can be accounted for by differences in
the mass accretion rate, with the Z sources probably accreting near
the Eddington rate and atoll sources at rates ranging from near
Eddington to less than 10\% of it, it is generally believed that 
additional differences, e.g. in the neutron star properties, are
required to explain {\it all} observational differences
\citep{hava1989}.

%The patterns traced out by Z sources resemble the letter 'Z' and
%consist of three branches, which from top to bottom are called the
%horizontal (HB), normal (NB) and flaring branch (FB); the patterns
%traced out by atoll sources generally consist of a banana shaped part
%(the banana state, usually divided into a lower (LB) and upper banana
%(UB)) and one or more fuzzy patches, connected to the lower banana,
%that are called islands or the island state (IS). 

The nature of \gx, one of the brightest neutron star LMXBs, is still
ambiguous. \citet{schatr1989} grouped it with the high luminosity
sources \citep{schatr1989}, which included the six sources that were
later labeled as Z sources. \citet{hava1989} classified it as a
bright atoll source, although they noted that of all atoll sources it
showed properties which were closest to those seen in the Z sources,
most notably its flaring branch-like appearance in the CD and the
featureless power law noise in the power spectrum that is typical for
that spectral state. Although the EXOSAT observations analyzed by
\citet{hava1989} only showed the source in the so-called 'banana
branch' state, the bimodal behavior of the source reported by
\citet{stwhta1985} strongly suggests that the source occasionally
enters a different state. The latter seemed to be confirmed by the
first observations of the source with the Rossi X-ray Timing Explorer
(RXTE), in which a clear two branched structure  was found in the CD
\citep{hovawi1998}. Although the pattern in the CD resembled both Z
and atoll source tracks, the resemblance to Z sources was
strengthened by the discovery (in the same observations) of a 57--69
Hz quasi-periodic oscillation (QPO), which had properties similar to
that of the horizontal branch QPO in the Z sources. The CD of GX 13+1
in a recent paper by \citet{murech2001}, which for the first time
displays a 'complete' pattern, shows a sharp vertex, similar to the
normal branch/flaring branch vertex seen in the Z sources and quite
unlike the rather smooth curves seen in atoll sources. More recently,
\citet{screva2003} analyzed a large set of RXTE data and conclude
that many of the source's properties do not fit within the atoll/Z
framework, although they favor the option of an atoll source. They
suggest that part of its unusual behavior can be explained with the
presence of a relativistic jet, that is almost pointing directly
towards us. 

Observations in the infrared suggest the presence of a K giant
secondary \citep{gagrba1992,bashch1999} and a possible orbital or
precessional modulation with a period of $\sim$20 days
\citep{bachsh2002}; these are properties that are thought to be more
typical for Z sources than for atoll sources.  \gx\ also shares a
common mean radio luminosity \citep{grse1986,gagrmo1988} with the Z
sources and black hole candidates \citep{fehe2000}, suggesting a
similar origin for the quiescent radio emission from persistent black
hole and Z source X-ray binaries, and \gx.  None of the (other) atoll
sources is consistent with this relation. Although the source is
variable in radio on time scales of less than an hour, no clear
relation between X-ray and radio luminosities was found
\citep{gagrmo1988}. 

In this paper we present the results of a coordinated X-ray/radio
campaign, which had as its principal aim to investigate the
possibility that \gx\ shows a similar X-ray state dependence of its
radio emission as is found in the Z sources. In most of those sources
the radio luminosity decreases from the horizontal branch to the
flaring branch. We find similar behavior in GX 13+1.

\section{Observations and analysis}\label{sec_obs}

\subsection{X-ray observations}\label{sec_obs_xray}

Our X-ray data were obtained simultaneously with the the Proportional
Counter Array \citep[PCA;][]{zhgija1993,jaswgi1996} and the High
Energy X-ray Timing Experiment
\citep[HEXTE;][]{grblhe1996,roblgr1998} onboard RXTE
\citep{brrosw1993}. The observations were performed in 1999 between
July 31 23:52 UTC and August 1 10:39 UTC (obs.\ 1), and on August 04
between 02:44 and 12:05 UTC (obs.\ 2). The total exposure times for
the two observations were, respectively, $\sim$22.5 ks and
$\sim$18.3 ks. The PCA and HEXTE data were obtained in several
different modes; the spectral and timing properties of these modes
are given in Table \ref{tab_modes}. For all modes we discarded data
taken during Earth occultations and passages through the South
Atlantic Anomaly.

\begin{table}[b]
\caption{Data modes for the RXTE/PCA and RXTE/HEXTE.}\label{tab_modes}
\begin{tabular}{lccc}
\hline
\hline
Mode & Time           & Energy      & Energy  \\
     & res.\ (s) & range (keV) & channels \\
\hline
{\tt Standard 1} & 1/8 & 2--60 & 1\\
{\tt Standard 2} & 16  & 2--60 & 129 \\
{\tt SB\_125us\_0\_13\_1s}  & 1/8192 & 2--5.9   & 1 \\
{\tt SB\_125us\_14\_17\_1s} & 1/8192 & 5.9--7.6 & 1 \\
{\tt E\_125us\_64M\_18\_1s} & 1/8192 & 7.6--60  & 64\\
\hline
{\tt E\_8us\_256\_DX1F} & 1/131072  & 10--250 & 256 \\
\hline
\end{tabular}
\end{table}

The {\tt Standard 2} data were used to produce light curves, colour
curves, a colour-colour diagram (CD), and a hardness-intensity
diagram (HID), and to perform a spectral analysis. Only data from
Proportional Counter Units (PCUs) 0 and 2 were used (all layers),
since these were the only two that were active during all our
observations. The data were background subtracted; dead time
corrections ($\sim$3--3.5\%) were only applied for the spectral
analysis. A soft colour was defined as the ratio of count rates in
the 4.2--7.5 keV and 2.5--4.2 keV energy bands, and a hard colour as
the ratio of count rates in the 10.0--18.5 keV and 7.5--10.0 keV
energy bands (these four energy bands correspond, respectively, to
{\tt Standard 2} channels (running from 1 to 129) 7--14, 3--6,
21--40, and 15--20).  A CD and a HID were produced by, respectively,
plotting hard colour versus soft colour and hard colour versus the
2.5--18.5 keV count rate, for each 16 s data point. The soft and hard
colour curves, which had an initial time resolution of 16 s, were
rebinned to a time resolution of 1024 s, to allow a better study of
their variations on long time scales. The PCA spectra were created
using the standard FTOOLS V5.2 routines. Systematic errors of 0.6\%
were added and response matrices were produced using PCARSP (V8.0).
The details of the models used to fit the spectra of \gx\ are
discussed in Section \ref{sec_res_xray}. 

\begin{figure*} \centering
\includegraphics[width=15.5cm]{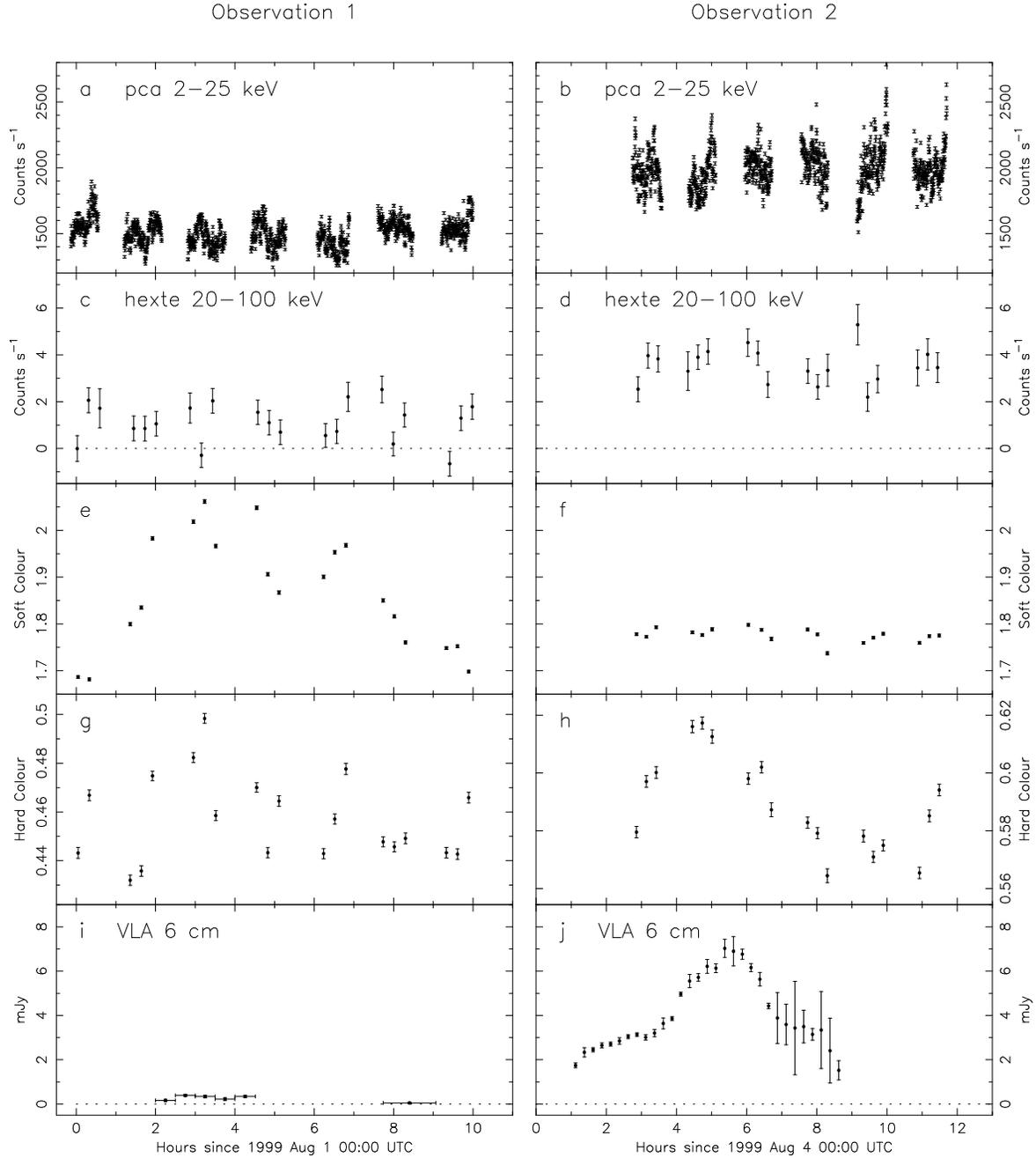} \caption{The
evolution of GX 13+1 during the observations on 1999 August 1 (left
column) and 1999 August 4 (right column). {\bf a,b}: PCA Count rate
in the 2--25 keV range (16 s bins). {\bf c,d}: HEXTE 20--100 keV
count rate (rebinned to 1024 s). {\bf e,f}: Soft color (rebinned to
1024 s). {\bf g,h}: Hard color (rebinned to 1024 s).  {\bf i,j}: VLA
6 cm radio flux density (1800--4800 s [i] and 900 s [j]). Errors on
the flux density in panel i are of the order of 50 $\mu$Jy. For the
PCA count rate and colours only data from PCUs 0 and 2 were used.
For the HEXTE count rate only Cluster A was used. Note that all
quantities have the same ranges for obs.~1 and 2, except for the hard
colour (to allow an easier comparison the with changes in the radio
flux density).} \label{fig_lcurves} \end{figure*}

HEXTE light curves were produced by running the FTOOLS V5.2 script
{\tt hxtlcurv} (which automatically corrects for background and
deadtime) on the {\tt E\_8us\_256\_DX1F} mode data of cluster A.
Additional rebinning was applied, resulting in 1024 s time bins, to
achieve better statistics.

The rapid X-ray time variability of the source was studied in terms
of power spectra. For this analysis we performed fast fourier
transforms (FFTs) of the high time resolution (1/8192 s) data . 
Power spectra were produced with frequencies of 1/1024--512 Hz
(2--26.5 keV band) and 1/16--2048 Hz (2--26.5 keV, 2--6.3 keV, and
6.3--26.5 keV bands). No background or dead time corrections were
applied to the data prior to the FFTs; the effects of dead time were
accounted for by our fitting method. The power spectra were selected
on time, count rate and/or colour and subsequently rms normalized
according to a procedure described in \citet{va1995b}. The resulting
power spectra were fitted with a constant, to account for the
dead-time modified Poisson level, a power law, a zero-centered
Lorentzian, to account for the deviations from the power law noise,
and a Lorentzian for a weak QPO in the second observation. Errors on
the fit parameters were determined using $\Delta\chi^2=1$. Upper
limits on QPOs were determined by fixing the frequency and FHWM to (a
range of) values and using $\Delta\chi^2=2.71$ (95\% confidence).

\begin{figure}[t] 
\resizebox{\hsize}{!}{\includegraphics{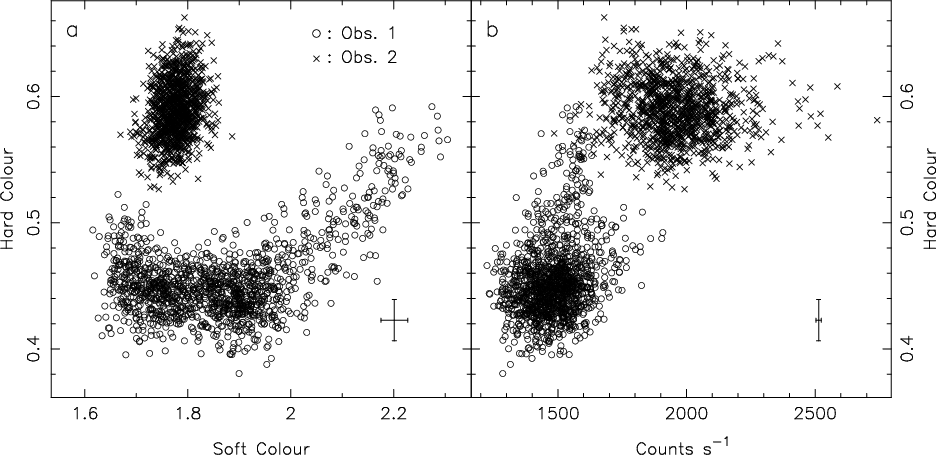}}
\caption{(a) Colour-colour diagram and (b) hardness-intensity diagram
for the two RXTE/PCA observations of GX 13+1. Points of the first
observations are depicted by the open circles, those of the second
observation by crosses. Each data point represent a 16 s interval.
Typical error bars are shown in the lower-right corners. 
}\label{fig_cd} \end{figure}

\subsection{Radio observations}\label{sec_obs_radio}

The radio data were obtained with the Very Large Array (VLA) radio
observatory in its most extended (A) configuration.. \gx\ was
observed at 6 cm on 1999 August 1 between  00:33 and 09:31 UTC (obs.\
1), and on 1999 August 4 between 00:22 and 08:51 UTC (obs.\ 2).
During the first observation 25 of the 27 VLA antennas were used; 26
were used during the second observation. Due to technical problems
$\sim$39\% (obs.\ 1) and $\sim$8\% (obs.\ 2) of the total observing
time was lost.  Observations were made simultaneously in both
circular polarizations in  each of two independent 50 MHz bands
centered on 4885.1 and 4835.1 MHz. Flux densities, which are all
Stokes I, were referenced to those of 1328+307 (3C 286), taken to be
7.462 and 7.510 Jy at 4885.1 and 4835.1 MHz respectively (Perley et
al., priv. comm.). The overall flux density scale is probably good to
at least $\sim5\%$. To calibrate the complex antenna gains the
standard calibration source 1817-254 was observed at 30 minute (obs.\
1) or 6 minute (obs.\ 2) intervals; the reason for the denser
sampling of the calibration source in obs.\ 2 was that during obs.\ 1
a large part of the data was rendered useless due to a combination of
bad weather conditions and too long intervals between calibration
observations. The data were reduced and analyzed using the
Astronomical Image Processing System (AIPS).  For each day, images
were first made using the entire data sets, ignoring any variability
of GX 13+1.  These images were used to find nearby confusing sources,
which were then subtracted from the original uv-data.  These uv-data
were then split into 15-minute bins (or longer, in the case of weak
signals) and imaged to give the flux density history of GX 13+1. This
procedure allows the full synthesis mapping of confusing sources,
while retaining the best possible sensitivity to fluctuations of GX
13+1. The flux densities are the average of the of maximum flux
density and the integrated flux density in the region containing the
source in the cleaned image. Note that because of calibration and
other "dead" time, the amount of data in the each bin is usually less
than the sampling time. Errors on the flux density were
conservatively calculated as the sum of the squares of (1) the
difference of the maximum and the integrated flux density and (2) the
measured off-source rms. Early in obs.~1 and during the beginning and
end of obs.~2 the source wandered away from the image center,
probably because the elevations of GX 13+1 and the calibrator were
rather different there - the effects of these problems can clearly be
seen in the increase of the error bars at the end of the radio light
cure in Fig.~\ref{fig_lcurves}j.

\section{Results}\label{sec_res}

\begin{figure}[t]
\resizebox{\hsize}{!}{\includegraphics{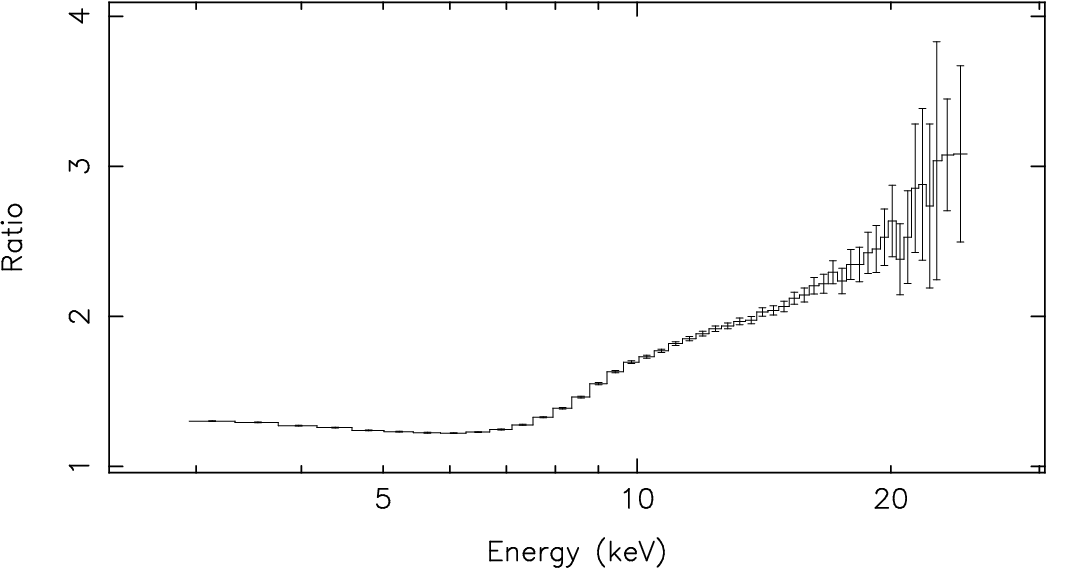}}
\caption{The ratio of the mean raw RXTE count rate spectrum of
observation 2 to that of observation 1, revealing an overall increase
and a relative hardening of the spectrum. }\label{fig_spec}
\end{figure}

\subsection{X-ray observations}\label{sec_res_xray}

From Figs.~\ref{fig_lcurves}--\ref{fig_spec} it is clear that the
X-ray properties of the source changed between the first to the
second observation. The second observation has a higher count rate,
both in the 2--25 keV (Fig.\ \ref{fig_lcurves}a and b) and 20--100
keV bands (Fig.\ \ref{fig_lcurves}c and d) and shows more variability
in the 2--25 keV band. Also, the second observation tends to be
spectrally harder; while the PCA count rates increased only by a
factor of $\sim$1.3, the HEXTE count rates increased by a factor of
$\sim$3.2. The spectral difference is most clearly seen in the CD and
HID shown in Fig.\ \ref{fig_cd}, where the two observations show up
as  distinct patches, and Fig.\ \ref{fig_spec}, which shows the ratio
of the spectra of obs.~2 and 1. Note that the small bridge between
the two patches in the HID (Fig.\ 2b) does not represent a real
connection between the two patches, but corresponds to the upper
right part of patch traced out during obs.\ 1 in the CD. It is not
clear from our observations how the source bridges the gap between
the two patches in HID - we refer to \citet{screva2003} for an
observation that shows a transition between the two patches. 
Although type I X-ray bursts have been observed from GX 13+1
\citep{fl1985,madomi1995}, none were seen during our observations.

%\begin{table}
%\caption{Spectral fit results. Errors on the fit parameters are 68\% confidence limits, upper limits represent 95\% confidence limits. The 3--10 and 10--25 keV fluxes are unabsorbed.}\label{tab_spectra}
%\begin{center}
%\begin{tabular}{lcc}
%\hline
%\hline
%Parameter			& Obs.\ 1			& Obs.\ 2 			\\
%\hline
%N$_H$ (atoms cm$^{-2}$) 	& \raisebox{.0ex}[2.5ex][.0ex]{$2.9\,10^{22}$} (fixed)	& $2.9\,10^{22}$ (fixed) 	\\
%kT$_W$ (keV)			& 0.88$\pm$0.01			& 0.84$\pm$0.05			\\
%kT$_e$ (keV)			& 2.84$\pm$0.08			& 2.58$\pm$0.05			\\
%$\tau$				& 7.4$\pm$0.2			& 9.8$\pm$0.3			\\
%R$_W$ (km)			& 19$\pm$4		        & 20$\pm$2			\\
%Photon index    		& 3.07 (fixed)			& 3.07$^{+0.30}_{-0.17}$	\\
%Power-law Norm.$^a$   		& $<$0.6			& 3.6$\pm$2.2			\\
%E$_{FE}$ (keV)  		& 6.42 (fixed)			& 6.42 (fixed)			\\
%FWHM (keV)			& 1.09$\pm$0.03			& 0.94$^{+0.20}_{-0.12}$	\\
%Fe EW (eV)			& 525				& 137				\\
%$F_{3-25}$  (ergs cm$^{-2}$s$^{-1}$) & $8.4\,10^{-9}$		& $1.15\,10^{-8}$		\\
%$F_{20-25}$ (ergs cm$^{-2}$s$^{-1}$) & $4.0\,10^{-11}$		& $1.12\,10^{-10}$		\\
%$\chi^2_{red}$ (dof) 		& 1.13 (42)			& 0.61 (41)			\\	
%$F$-test			& ---$^b$				& $6.7\,10^{-14}$		\\
%\hline
%\end{tabular}
%\end{center}
%\noindent $^a$ Normalization in units of photons\,keV$^{-1}$\,cm$^{-2}$\,s$^{-1}$ \\
%\noindent $^b$ No $F$-test possible due to increase of $\chi^2_{red}$
%\end{table}

\subsubsection{X-ray spectra}

X-ray spectral fits were performed with XSPEC \citep[V11.2]{ar1996}
to the 3-25 keV PCA spectra using two different models. The first
model is based on the continuum model used by \citet{ueasya2001}, for
their ASCA 1--10 keV spectra of \gx, and \citet{sipaoo2002}, for
their 2--10 keV XMM-Newton spectra. It consists of a black body ({\tt
bbody} in XSPEC), and a disk black body ({\tt diskbb}). Galactic
neutral hydrogen absorption ({\tt phabs}) was fixed at
$N_H=2.9\,10^{22}$ atoms cm$^{-2}$ \citep{ueasya2001}. A reasonable
fit was found for obs.~2 ($\chi^2_{red}=1.52$ - with an unphysically
small disk radius), but not for obs.~1 ($\chi^2_{red}=11.3$). Adding
a line ({\tt gauss}) around 6.4 keV and an absorption edge ({\tt
edge}) around 8 keV \citep{ueasya2001} did not lead to acceptable
fits.  The second model we tried was a thermal Comptonization model
\citep[{\tt comptt},][]{ti1994,huti1995}, which has recently been
applied successfully to the broad band continuum of several bright
neutron star LMXBs \citep[see, e.g.,][]{distro2000,iabudi2001}. The
$N_H$ was again fixed to $N_H=2.9\,10^{22}$ atoms cm$^{-2}$. For the
second observation a reasonable fit was obtained
($\chi^2_{red}$=1.35), but the first observation showed large
residuals between 5 and 10 keV ($\chi^2_{red}$=15.6). Adding a line 
around 6.4 keV and an edge around 9 keV greatly improved the fit
($\chi^2_{red}$=0.6). For the second observation the addition of the
line and edge led to a small improvement  ($\chi^2_{red}$=0.9), with
the two components not being detected significantly. The best-fit
results with the four-component model are given in Table
\ref{tab_spectra}. Some caution should be taken with interpreting the
spectral results, as the low line energy (which tended to decrease
when not fixed) and the high edge energy indicate different
ionization stages of Fe. Also the large spectral changes within
obs.~1 may have been partly responsible for the observed residuals.
For both observations no black body component was needed, as the
Comptonization component provided good fits at low energies. 

\begin{table}[t]
\caption{Spectral fit results. Errors on the fit parameters are 68\% confidence limits, upper limits represent 95\% confidence limits. The 3--10 and 10--25 keV fluxes are unabsorbed.}\label{tab_spectra}
\begin{center}
\begin{tabular}{lcc}
\hline
\hline
Parameter			& Obs.\ 1			& Obs.\ 2 			\\
\hline
N$_H$ (atoms cm$^{-2}$) 	& \raisebox{.0ex}[2.5ex][.0ex]{$2.9\,10^{22}$} (fixed)	& $2.9\,10^{22}$ (fixed) 	\\
kT$_W$ (keV)			& 0.98$\pm$0.02			& 0.92$\pm$0.01			\\
kT$_e$ (keV)			& 2.88$\pm$0.06			& 3.04$\pm$0.04			\\
$\tau$			& 7.2$\pm$0.2			& 8.10$\pm$0.15			\\
R$_W$ (km)			& 21.7$\pm$0.9		        & 27.2$\pm$0.7			\\
E$_{FE}$ (keV)  		& 6.42 (fixed)			& 6.42 (fixed)			\\
FWHM (keV)			& 0.78$\pm$0.13			& 0.4$^{+0.2}_{-0.4}$	\\
Fe EW (eV)			& 223				& $<$96				\\
$E_{edge}$ (keV)                & 9.1$\pm$0.1                   & 9.1 (fixed)	\\
$\tau_{edge}$                   & 0.12$\pm$0.03                 & $<$0.03 \\ 
$F_{3-10}$  (ergs cm$^{-2}$s$^{-1}$) & $1.5\,10^{-8}$		& $2.0\,10^{-8}$		\\
$F_{10-25}$ (ergs cm$^{-2}$s$^{-1}$) & $1.8\,10^{-9}$		& $3.6\,10^{-9}$		\\
$\chi^2_{red}$ (dof) 		& 0.6 (40)			& 0.9 (41)			\\	
\hline
\end{tabular}
\end{center}
\noindent Fit parameters of the {\tt comptt} model - kT$_W$: input soft photon (Wien) temperature - kT$_e$: plasma temperature - $\tau$:  plasma optical depth - R$_W$: effective Wien radius for the soft seed photons \citep{invest1999}, here derived for an assumed distance of 7 kpc
\end{table}

For a more model independent comparison of the spectra of the two
observations we plot the ratio of observations 2 and 1 in Fig.~\ref{fig_spec}. While below 7
keV the ratio is rather constant, above that energy the spectrum of
obs.~2 becomes increasingly hard.

\subsubsection{X-ray variability}

\begin{figure}[t]
\resizebox{\hsize}{!}{\includegraphics{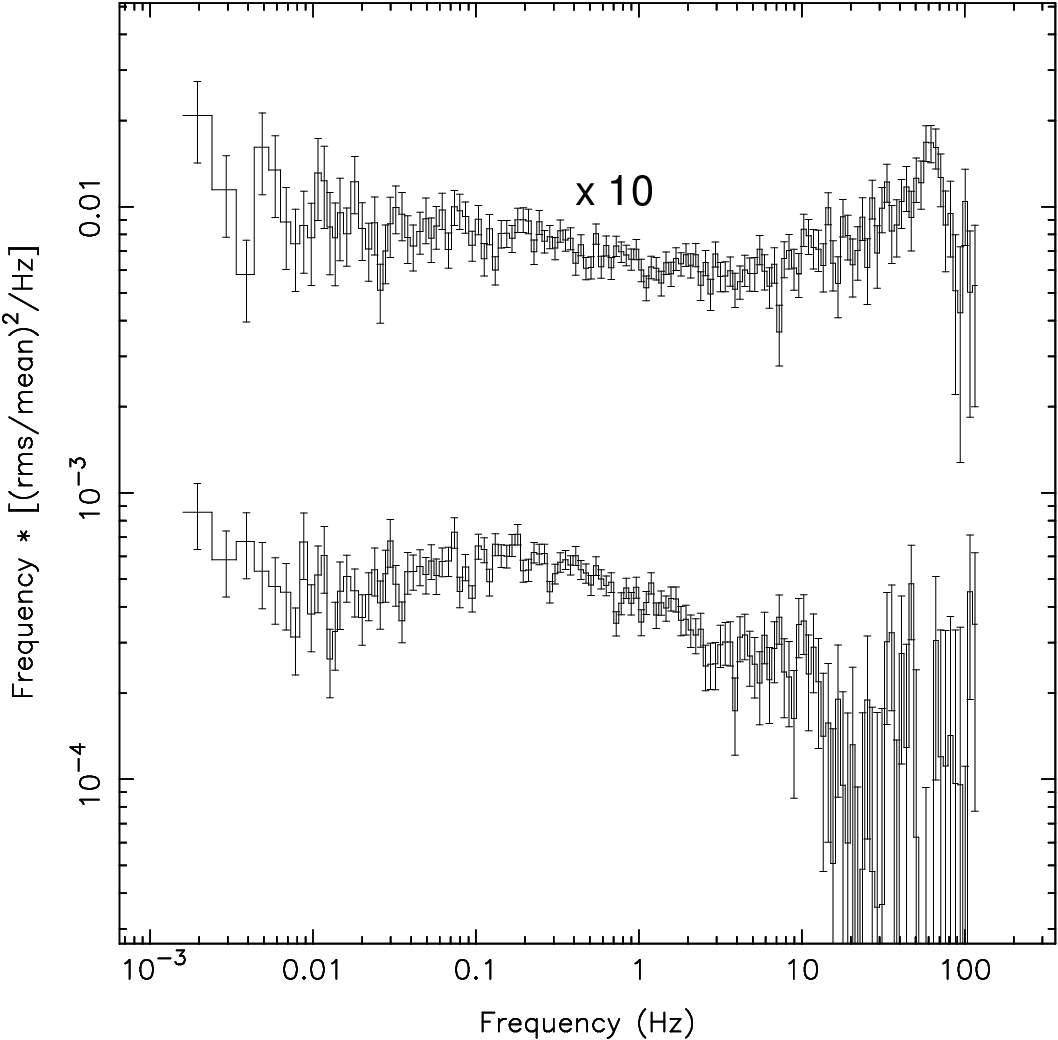}}
\caption{Power spectra (2--26.5 keV) of the two RXTE/PCA observations (lower: obs.\
1, upper: obs.\ 2) in an $\nu P(\nu)$ representation. Note that the
upper power spectrum has been multiplied by a factor of 10, to avoid
overlapping of the power spectra. The Poisson level has in both cases
been subtracted for plotting purposes.}\label{fig_pds}  \end{figure}

From the PCA light curves in Fig.~\ref{fig_lcurves} it is clear that,
at least on time scales of minutes to hours, the source was more
variable during the second observation. The 2--26.5 keV power spectra
of the two observations are shown in Fig.\ \ref{fig_pds}; the results
of the power spectral fits are given in Table \ref{tab_pds}. Both
power spectra are relatively featureless and their continuum can be
well fit by a combination of power law noise and a zero-centered
Lorentzian. While the power law noise has an identical index in the
two observations, the FHWM of the two zero-centered Lorentzians, and
hence their maxima in a $\nu P_{\nu}$ plot, differ by more than two
orders of magnitude, suggesting that they might not be related. We
added a Lorentzian to the fit function of the second observation, for
a QPO at 60 Hz - this QPO has a statistical significance of
3.2$\sigma$. No similar QPO was found in the first observation. We
also measured the total rms in two frequency intervals; 0.001--1 Hz
and 1--100 Hz. In both frequency intervals the fractional variability
is stronger in the second observations, with the difference being
more pronounced at high frequencies. We searched for high frequency
QPOs in the 1/16--2048 Hz power spectra with the FWHM fixed to 150
Hz, but only upper limits could be determined. For obs.\ 1 they were:
2.1\% (2--26.5 keV), 2.8\%(2--6.3 keV), and 4.0\% (6.3--26.5 keV) and
those for obs.\ 2: 2.2\% (2--26.5 keV) , 3.7\% (2--6.3 keV), and
3.6\% (6.3--26.5 keV).

\subsection{Radio observations}

During the first observation the source was only sometimes detected
significantly, at a very low level and with low variability; the
average flux density was $\sim$0.25 mJy. The radio flux density in
the second observations was much higher, varying between 1.3 and 7.2
mJy. The flux density showed a slow increase for about 4.5 hours,
followed by a gradual decrease. No significant variability was
detected on time scales less than 15 minutes. An upper limit on the size of the source of 0.15 arcsecond was obtained.

\subsection{X-ray/radio comparison}

Comparing panels i and j of Fig.\ \ref{fig_lcurves} with panels a and
b of the same figure shows that, apart from the fact that the second
observation is brighter both in X-rays and radio, there is no
apparent relation between X-ray count rate and radio flux, at least
not on the time scales on which the latter varies. On the other hand,
a comparison of the radio flux and hard X-ray colour of obs.\ 2
reveals similar broad peaks centered around $\sim$05:00 UTC. As is
obvious from panels \ref{fig_lcurves}h and \ref{fig_lcurves}j there
exists a significant lag between the peaks of the hard colour curve
and the radio light curve. Fitting the broad peaks in the hard color (02:52--06:02 UTC) and radio curves (03:53--06:53 UTC) both with a Lorentzian gives a lag between the two peaks of 42$\pm$3 minutes.

%Note that the peak in the hard colour curve seems to be slightly
%broader than the one in the radio light curve. 

\begin{table}[t]
\begin{center}
\caption{Power spectral fit results. As a model independent 
measure of the variability we also include the total fractional rms in the
0.001--1 Hz and 1--100 Hz ranges. FWHM stands for full-width-at-half-maximum.}\label{tab_pds}
\begin{tabular}{lcc}
\hline
\hline
Parameter                  & Obs.\ 1 & Obs.\ 2 \\
\hline
Power law rms$^a$ (\%)     & 5.15$\pm$0.12	     & 7.71$\pm$0.11 \\
Power law index            & 1.09$\pm$0.01	     & 1.09$\pm$0.01 \\
Lorentzian rms$^b$ (\%)    & 2.99$\pm$0.01           & 3.4$\pm$0.6 \\
Lorentzian FWHM (\%)       & 0.45$\pm$0.04           & 100$\pm$30 \\
Lorentzian freq. (Hz)      & 0 {\it (fixed)}         & 0 {\it (fixed)} \\
QPO rms$^c$ (\%)           & $<$1                    & 2.1$\pm$0.4 (3.2$\sigma$)$^d$ \\
QPO FHWM (Hz)              & 18 {\it (fixed)}        & 18$\pm$6 \\
QPO frequency (Hz)         & 60 {\it (fixed)}        & 60$\pm$2 \\
rms [0.001--1 Hz] (\%)     & 6.6$\pm$0.02            & 9.15$\pm$0.02 \\
rms [1--100 Hz]   (\%)     & 3.17$\pm$0.10           & 5.93$\pm$0.06 \\
$\chi^2_{red} (d.o.f)$       & 1.07 (209)                & 1.01 (210)  \\
\hline
\end{tabular}
\end{center}
\noindent $^a$ Integrated between 0.001 and 1 Hz 
\noindent $^b$ Integrated between 0 and $+\infty$ Hz 
\noindent $^c$ Integrated between $-\infty$ and $+\infty$ Hz  
\noindent $^d$ Significance is calculated from the power, not the fractional r.m.s.  
\end{table}

\section{Discussion}\label{sec_disc}

We have observed the neutron star LMXB \gx\ simultaneously in X-rays
and radio. The source was found in two clearly distinct X-ray states;
the spectrally hard state had associated radio fluxes that were
between 4 and 18 time higher than the maximum we detected in the
softer X-ray state. This dependence of the radio flux on X-ray state
is similar to what is found for other bright neutron star LMXBs
\citep[the Z sources; see][and references therein]{hjha1995} and more
recently also in the less-luminous atoll source 4U 1728-34
\citep{miferu2003}. More specifically, our observations strongly
suggest that the radio flux of the source is related to the X-ray
spectral hardness, on time scales of both hours and days.

\subsection{Outflow}

It is generally believed that the radio emission from neutron star
and black hole X-ray binaries is produced by highly relativistic
electrons that interact with magnetic fields to produce synchrotron
radiation. High resolution radio observations of a handful of X-ray
binaries show that these electrons reside in powerful, collimated
outflows, commonly referred to as jets. It is assumed that in the
X-ray binaries for which jets have not (yet) been directly observed
the radio emission originates in a similar outflow. To see whether
this could also be the case for GX 13+1, we estimate the size of the
emission region. If, for an assumed spherical region, this size is
larger than the binary separation, the highly relativistic
synchrotron plasma cannot be contained in the system and the most
plausible option would then be that we are dealing with an outflow.
Assuming a maximum brightness temperature $T_B$ of $\leq10^{12}$ K
\citep[see e.g.][]{kepa1969} and a distance of 7 kpc
\citep{bashch1999} we derive, following \citet{fe2003}, a minimum
size for the emitting region of $\sim$15 $R_\odot$ ($10^{12}$ cm) at
the peak of the radio flare in obs.~2. A size estimate based on the
fastest part (1.5 hr) of the rise of the radio flare ($\leq c\Delta
t\sim1.6\cdot10^{14}$ cm) is consistent with this value. Even for a
system with equal masses for the primary and secondary  and an
orbital period of 20 days the emission region is at least comparable
and probably larger than the size of the binary system
($\sim3\cdot10^{12}$ cm), suggesting an outflow is also present in GX
13+1.

\subsection{X-ray/radio connection}

Although the exact mechanism for producing jet outflows in X-ray
binaries is still not clear, the energy needed to achieve the inferred
observed high bulk velocities suggests that they form in the inner
parts of the accretion flow, where also most of the X-rays are
produced. Simultaneous X-ray and radio observations of GRS 1915+105
seem to confirm such a direct link between the outflow and inner
accretion disk \citep{midhch1998,klfepo2002}. Clear patterns of
X-ray/radio behavior are also observed in other black hole and Z
source X-ray binaries; in general the spectrally hard X-ray states
have a higher radio luminosity than the spectrally soft states.

Our observations of \gx\ reveal a similar pattern, with the second
observation being both more luminous in the radio and showing a
harder energy spectrum (Figs.~\ref{fig_lcurves}--\ref{fig_spec}).
While the change between the soft and hard spectral states of \gx\
most likely occurred on a time scale of one or two days, we also find
a relation between the X-ray and radio properties on a time scale of
a few hours: a hardening of the X-ray spectrum was followed by an
increase in the radio luminosity with a delay of $\sim$40 minutes.
This is the first time such a short-term X-ray/radio connection and
delay have been found and measured in a neutron star LMXB. It is
interesting to note that a similar delay has also been found on
several occasions in the galactic black hole X-ray binary GRS
1915+105 \citep{klfepo2002}. In that case however, the delay
($\sim$45--60 minutes) was with respect to the start of the dips in
the RXTE/PCA count rate. These dips coincided with a strong spectral
hardening. Also, the X-ray dips and radio flares in GRS 1915+105 were
a rapid recurring phenomenon, whereas the radio flare in GX 13+1 was 
more likely a singular event. \citet{klfepo2002} explained the
observed delay in GRS 1915+105 as the time it takes for the flow to
become optically thin in the radio; this effect is clearly observed
in Fig. 9 of \citet{dhmiro2000}, which shows light curves of the
directly imaged jet of GRS 1915+105 at different radio wavelengths
that peak later with increasing wavelength \citep[see
also][]{pofe1997}. Observations of Sco X-1 \citep{fogebr2001} suggest
that such a delay can also result from the transfer time of the
energy from the core to the radio lobes. 

%The fact that, unlike in GRS 1915+105, no X-ray dip can be associated
%with the radio flare GX 13+1 could possibly be due to the different
%nature of the compact object; if the inner disc flow changes to a
%radiatively inefficient flow (as has been suggested by e.g.\
%\citet{bemeki1997} to explain the dips in GRS 1915+105) the effect of
%this is expected to be much more pronounced for black holes than for
%neutron stars. 

\begin{table}[t]
\caption{Observed mean and maximum radio flux densities and estimated distances for the six Z sources and GX 13+1. This table is reproduced from Tables 2 and 3 in \citep{fehe2000}, with updated values for GX 13+1.}\label{tab:radio}
\begin{center}
\begin{tabular}{lcccc}
\hline
Source & Mean$^a$        & Max$^b$      & distance & Refs\\
       & (mJy)           &  (mJy)       & (kpc)    & \\
\hline
Sco X-1  & $10 \pm 3$    & 22 & $2.0 \pm 1.0$ & 1,2 \\
GX 17+2  & $1.0 \pm 0.3$ & 13.4 & $7.5 \pm 2.3$ & 1,3\\
GX 349+2 & $0.6 \pm 0.3$ & 1.3 & $5.0 \pm 1.5$ & 4,5\\
Cyg X-2  & $0.6 \pm 0.2$ & 3.4 & $8.0 \pm 2.4$ & 1,6,7\\
GX 5-1   & $1.3 \pm 0.3$ & 1.6 & $9.2 \pm 2.7$ & 1\\
GX 340+0 & $0.6 \pm 0.3$ & 0.6 & $11.0 \pm 3.3$ & 8\\
GX 13+1  & $1.8 \pm 0.3$ & 7.7 & $7 \pm 1$ & 9--13 \\
\hline
\end{tabular}
\end{center}

\noindent $^a$ Mean cm radio flux density\\
\noindent $^b$ Maximum radio flux density at 6 cm

\vspace{0.1cm}
Refs 
1: \citet{pe1989},
2: \citet{brfoge1999}
3: \citet{pelezi1988},
4: \citet{copo1991},
5: \citet{chsw1997},
6: \citet{hjhaco1990},
7: \citet{cocrhu1979},
8: \citet{pezwva1993},
9: \citet{grse1986},
10: \citet{gagrmo1988},
11: \citet{befeku1999},
12: \citet{bashch1999},
13: this work
\end{table}

The ratio of (radio flux density change)/(X-ray flux change) is much
larger in GX 13+1 than in GRS 1915+105. If we assume that the inner
disk is ejected to form the outflow, as has been proposed for GRS
1915+105 to explain the X-ray variations, our second observation of
GX 13+1 suggests  that a smaller fraction of the inner disk is
ejected in this case. The fact that clear dips are observed in GRS
1915+105 might be related to the different nature of the compact
object, or to a much larger amount of matter being expelled.
Following \citet{fe2003} we can estimate the mean power of the radio
event in obs.~2 to be $\sim1.8\cdot10^{36}$ erg s$^{-1}$. While this
is only $\sim$1 percent of the 3--25 keV X-ray luminosity
($\sim1.4\cdot10^{38}$ erg s$^{-1}$), the kinetic energy associated
to the (possibly relativistic) outflow may increase this number to a
larger fraction of the total accretion energy.

Although short term variations in the radio luminosity of \gx\ have
been found before, they were only compared to the X-ray count rates
\citep{gagrmo1988}, and not to the X-ray spectral properties. No
correlations between radio luminosity and X-ray count rate were
found, most likely because X-ray data were only available for the
radio weak part of their data set. The time scale of shortest radio
fluctuations observed by \citet{gagrmo1988} (a few hours) is
consistent with that of the flare in our second observation. The 6 cm
peak flux density they measured was 2.2 mJy, about a factor of
$\sim$3.3 lower than our maximum flux density. Interestingly, when
fitting our X-ray spectra with the same $N_H$ used by
\citet{gagrmo1988} and using their assumed distance of 7 kpc we find
a similar difference for the 1--20 keV luminosity.

\subsection{Comparison with Z and atoll source observations}

Before discussing the nature of GX 13+1 in view of the atoll/Z source
classification we compare its X-ray/radio behavior with that of
sources from both classes. Although \citet{screva2003} concluded that
GX 13+1 is neither a Z nor an atoll source, one could, purely based
on its appearance in the CD, argue that the source was in the atoll
source banana branch and island state, during obs.~1 and 2
respectively, or the Z source flaring branch and normal branch.

Based on observing campaigns of the Z sources GX 17+2
\citep{pelezi1988}, Cyg X-2 \citep{hjhaco1990}, Sco X-1
\citep{hjstwi1990} and GX 5-1 \citep{talehj1992}, \citet{pe1989}
suggested that all Z sources share a common radio, UV, and  X-ray
luminosity on their normal branch. Subsequent detections of GX 349+2
\citep{copo1991} and GX 340+0 \citep{pezwva1993} at approximately the
predicted radio brightness seemed to confirm this idea.
\citet{pe1989} derived a normal branch luminosity of
$\sim1.6\cdot10^{38}$ erg s$^{-1}$ in the 1.5-15 keV band. In the
same energy band we obtain (extrapolating our fit to lower energies)
$\sim1.7\cdot10^{38}$ erg s$^{-1}$, which is remarkably close the
value of \citet{pe1989}. 

The assumption of \citet{pe1989} might be not completely valid; using
the distance estimates and average flux densities from
\citet{fehe2000} (see Table~\ref{tab:radio}) it seems that there is a
considerable spread in the average radio luminosity of  the six Z
sources. Moreover, none of them has a higher value than GX 13+1.
However, this average radio luminosity depends strongly on the time a
source spends in a radio bright state. Taking the maximum radio flux
densities (at 6 cm) reported in the literature (see references in
Table~\ref{tab:radio}) should partly correct for this, which results
in GX 17+2 being the most luminous with GX 13+1 being second. The
overall behavior of the radio brightness along the track in the CD is
also similar to that observed in several of the Z sources
\citep{pelezi1988, hjhaco1990, hjstwi1990}.

The only atoll source for which a clear pattern in the X-ray/radio
emission has been found is 4U 1728--34 \citep{miferu2003}. In that
source the radio flux density was highest in the island state; it
increased by a factor $\sim$6 from the hard part of the island state
toward the transition between the island state and, what seemed to
be, the (softer) banana branch. After this transition it  dropped
sharply by a factor of $\sim$6. While the decrease of radio flux with
spectral hardness (in the island state) is opposite to overall
behavior in GX 13+1 and several Z sources (where we see an increase
with spectral hardness along the track in the CD), the factor of
$\sim$6 difference between the island state and banana branch is
similar in sign and magnitude to the difference between  our second
and   first observation. The average X-ray luminosity of 4U 1728--34
(assuming a distance of 5.2 kpc \citep{gapsch2002}) is more than a
factor 10 lower than that of GX 13+1, whereas the 6 cm flux density
is probably more than a factor 100 lower.

\subsection{Z or atoll?}

The nature of GX 13+1 and its place in the Z/atoll classification
scheme have recently been extensively studied and discussed by
\citet{screva2003}. They found that its motion through the HID is in
the opposite sense to that in the CD, with the X-ray count rate
increasing again when the source moves into the spectrally hard
state, contrary to most atoll and Z sources (see our
Fig.~\ref{fig_cd} and also \citet{wivaku1997} for similar behavior in
the Z source Cyg X-2). Moreover, the strength of the very low
frequency variability also changes in the opposite sense to that
observed in Z and atoll sources. 

In an attempt to fit GX 13+1 within the Z/atoll scheme as an atoll
source,  \citet{screva2003} tried to explain the source's peculiar
behavior with the presence of a relativistic jet with an axis nearly
aligned to our line of sight. The radio emission could then be
boosted by more than an order of magnitude, putting the radio
emission of GX 13+1 in accordance with measurement and upper limits
of other atoll source; the unusual X-ray phenomena could be the
consequence of a better view of the X-ray emitting regions associated
with the base of the jet, possibly assisted by Doppler boosting. 

As we showed above, the X-ray and radio luminosities of GX 13+1 are
actually in the range expected for Z sources,  without requiring any
unusual jet geometries, making a Z source nature more likely in our
opinion (the atoll sources GX 9+1, GX 9+9 and GX 3+1 have also
similar X-ray luminosities but much lower radio luminosities). This
is supported by our spectral fit parameters, which are quite similar
to those found in the Z sources \citep[see
e.g.][]{distro2000,diroia2001} and less like those in the atoll
sources \citep[see e.g.][]{diiabu2000,baolbo2000,baol2002}, and by
the fact that GX 13+1 rarely shows type I X-ray bursts. Based on the
morphology in our CD GX 13+1 was probably observed in the normal
branch and flaring branch. Although no clear indications for normal
branch oscillations are found, the band limited noise measured by
\citet{screva2003} has properties similar to the normal branch
oscillations found in GX 5-1 \citep{jovaho2002} and GX 340+0
\citep{jovawi2000} close to the normal branch/flaring branch vertex,
where it is very broad ($Q<1$) and rather weak ($\le$2\% rms). In GX
5-1 and GX 340+0 this broad bump evolves into a strong 6 Hz QPO as
the spectrum hardens - this is not observed in GX 13+1. There are
additional differences with the other Z sources, in particular with
respect to the behavior of the low frequency variability. 

Finally, we note that although the source was more similar to the Z sources than to the atoll sources during our observations (at a luminosity three times as high as during the EXOSAT era), this does not exclude that at lower luminosities it behaves more atoll-like.

%As mentioned before, based on the data that they analyzed 
%\citet{hava1989} originally classified \gx\ as an atoll source, even
%though it showed similarities to the Z sources in their FB. The main
%arguments for their choise were the relatively flat slope of the
%power law noise, the presence of wiggles around the power law noise
%(also visible in the power spectrum of obs.\ 1, see Fig.\
%\ref{fig_pds}), and the lack of strong band limited noise. Our data,
%which seem to cover a wider range of X-ray states than the data used
%by \citet{hava1989}, show that the atoll classification is probably
%no longer justified. Based on its appearance in the CD one could
%argue that the observed track resembles that of several other atoll
%sources, with the source being in the banana state during obs.\ 1 and
%the island state during obs.\ 2. However, the CDs of \gx\ found in
%\citet{murech2001} and Schnerr et al. (2002, in prep.) show ...

%We also note that for sources with such a 'complete' track the ratio %of fmin and fmax (Muno et al.) is much closer to that of the Z %sources. Also infra red %

%The CD in Fig.\ \ref{fig_cd} seems to ... to the standard picture of
%an atoll source CD. It shows shows a banana branch, that can be
%divided in an upper and a lower part, and connected to its lower part
%an island state. 

%Problems: 1) The count rate in the island state is higher than that %in the banana state. 2) The band limited noise in the 

\section{Summary} Our simultaneous X-ray/radio observations of GX
13+1 revealed a strong dependence of the radio brightness on the
X-ray state of the source. In the hard spectral state, which was at
least 4 times brighter in the radio than the soft state, we also
found a correlation between the X-ray and radio properties on a short
time scale, with changes in radio having a delay of $\sim$40 minutes
with respect to those in X-rays. We attribute this delay to the time
it takes for the flow to become optically thin in the radio. The
absence of strong dips in the X-ray light curve during the radio
flare suggests that only a small amount of the matter is redirected
from the inflow to the outflow. On the basis of a comparison with
atoll and Z sources we conclude that the source is more similar to
the Z sources, although several properties of GX 13+1 remain
unexplained.

\begin{acknowledgements} JH thanks Roald Schnerr for comments on an
earlier version of the manuscript and Jon Miller for his help with
the spectral data reduction. JH also acknowledges support from
Cofin-2000 grant MM02C71842. The National Radio Astronomy Observatory
is a facility of the National Science Foundation operated under
cooperative agreement by Associated Universities, Inc.

\end{acknowledgements}

%\bibliographystyle{aa}
%\bibliography{all-bib}

\begin{thebibliography}{57}
\expandafter\ifx\csname natexlab\endcsname\relax\def\natexlab#1{#1}\fi

\bibitem[{{Arnaud}(1996)}]{ar1996}
{Arnaud}, K.~A. 1996, in ASP Conf. Ser. 101: Astronomical Data Analysis
  Software and Systems V, Vol.~5, 17

\bibitem[{{Bandyopadhyay} {et~al.}(2002){Bandyopadhyay}, {Charles}, {Shahbaz},
  \& {Wagner}}]{bachsh2002}
{Bandyopadhyay}, R.~M., {Charles}, P.~A., {Shahbaz}, T., \& {Wagner}, R.~M.
  2002, \apj, 570, 793

\bibitem[{{Bandyopadhyay} {et~al.}(1999){Bandyopadhyay}, {Shahbaz}, {Charles},
  \& {Naylor}}]{bashch1999}
{Bandyopadhyay}, R.~M., {Shahbaz}, T., {Charles}, P.~A., \& {Naylor}, T. 1999,
  \mnras, 306, 417

\bibitem[{{Barret} \& {Olive}(2002)}]{baol2002}
{Barret}, D. \& {Olive}, J. 2002, \apj, 576, 391

\bibitem[{{Barret} {et~al.}(2000){Barret}, {Olive}, {Boirin}, {Done},
  {Skinner}, \& {Grindlay}}]{baolbo2000}
{Barret}, D., {Olive}, J.~F., {Boirin}, L., {et~al.} 2000, \apj, 533, 329

\bibitem[{{Berendsen} {et~al.}(2000){Berendsen}, {Fender}, {Kuulkers}, {Heise},
  \& {van der Klis}}]{befeku1999}
{Berendsen}, S.~G.~H., {Fender}, R., {Kuulkers}, E., {Heise}, J., \& {van der
  Klis}, M. 2000, \mnras, 318, 599

\bibitem[{{Bradshaw} {et~al.}(1999){Bradshaw}, {Fomalont}, \&
  {Geldzahler}}]{brfoge1999}
{Bradshaw}, C.~F., {Fomalont}, E.~B., \& {Geldzahler}, B.~J. 1999, \apjl, 512,
  L121

\bibitem[{{Bradt} {et~al.}(1993){Bradt}, {Rothschild}, \& {Swank}}]{brrosw1993}
{Bradt}, H.~V., {Rothschild}, R.~E., \& {Swank}, J.~H. 1993, \aaps, 97, 355

\bibitem[{{Christian} \& {Swank}(1997)}]{chsw1997}
{Christian}, D.~J. \& {Swank}, J.~H. 1997, \apjs, 109, 177

\bibitem[{{Cooke} \& {Ponman}(1991)}]{copo1991}
{Cooke}, B.~A. \& {Ponman}, T.~J. 1991, \aap, 244, 358

\bibitem[{{Cowley} {et~al.}(1979){Cowley}, {Crampton}, \&
  {Hutchings}}]{cocrhu1979}
{Cowley}, A.~P., {Crampton}, D., \& {Hutchings}, J.~B. 1979, \apj, 231, 539

\bibitem[{{Dhawan} {et~al.}(2000){Dhawan}, {Mirabel}, \&
  {Rodr{\'{\i}}guez}}]{dhmiro2000}
{Dhawan}, V., {Mirabel}, I.~F., \& {Rodr{\'{\i}}guez}, L.~F. 2000, \apj, 543,
  373

\bibitem[{{Di Salvo} {et~al.}(2000){Di Salvo}, {Iaria}, {Burderi}, \&
  {Robba}}]{diiabu2000}
{Di Salvo}, T., {Iaria}, R., {Burderi}, L., \& {Robba}, N.~R. 2000, \apj, 542,
  1034

\bibitem[{{Di Salvo} {et~al.}(2001){Di Salvo}, {Robba}, {Iaria}, {Stella},
  {Burderi}, \& {Israel}}]{diroia2001}
{Di Salvo}, T., {Robba}, N.~R., {Iaria}, R., {et~al.} 2001, \apj, 554, 49

\bibitem[{{di Salvo} {et~al.}(2000){di Salvo}, {Stella}, {Robba}, {van der
  Klis}, {Burderi}, {Israel}, {Homan}, {Campana}, {Frontera}, \&
  {Parmar}}]{distro2000}
{di Salvo}, T., {Stella}, L., {Robba}, N.~R., {et~al.} 2000, \apjl, 544, L119

\bibitem[{{Fender}(2003)}]{fe2003}
{Fender}, R. 2003, astro-ph/0303339

\bibitem[{{Fender} \& {Hendry}(2000)}]{fehe2000}
{Fender}, R.~P. \& {Hendry}, M.~A. 2000, \mnras, 317, 1

\bibitem[{{Fleischman}(1985)}]{fl1985}
{Fleischman}, J.~R. 1985, \aap, 153, 106

\bibitem[{{Fomalont} {et~al.}(2001){Fomalont}, {Geldzahler}, \&
  {Bradshaw}}]{fogebr2001}
{Fomalont}, E.~B., {Geldzahler}, B.~J., \& {Bradshaw}, C.~F. 2001, \apjl, 553,
  L27

\bibitem[{{Galloway} {et~al.}(2002){Galloway}, {Psaltis}, {Chakrabarty}, \&
  {Muno}}]{gapsch2002}
{Galloway}, D.~K., {Psaltis}, D., {Chakrabarty}, D., \& {Muno}, M.~P. 2002,
  astro-ph/0208464, 8464

\bibitem[{{Garcia} {et~al.}(1992){Garcia}, {Grindlay}, {Bailyn}, {Pipher},
  {Shure}, \& {Woodward}}]{gagrba1992}
{Garcia}, M.~R., {Grindlay}, J.~E., {Bailyn}, C.~D., {et~al.} 1992, \aj, 103,
  1325

\bibitem[{{Garcia} {et~al.}(1988){Garcia}, {Grindlay}, {Molnar}, {Stella},
  {White}, \& {Seaquist}}]{gagrmo1988}
{Garcia}, M.~R., {Grindlay}, J.~E., {Molnar}, L.~A., {et~al.} 1988, \apj, 328,
  552

\bibitem[{{Grindlay} \& {Seaquist}(1986)}]{grse1986}
{Grindlay}, J.~E. \& {Seaquist}, E.~R. 1986, \apj, 310, 172

\bibitem[{{Gruber} {et~al.}(1996){Gruber}, {Blanco}, {Heindl}, {Pelling},
  {Rothschild}, \& {Hink}}]{grblhe1996}
{Gruber}, D.~E., {Blanco}, P.~R., {Heindl}, W.~A., {et~al.} 1996, \aaps, 120,
  C641

\bibitem[{{Hasinger} \& {van der Klis}(1989)}]{hava1989}
{Hasinger}, G. \& {van der Klis}, M. 1989, \aap, 225, 79

\bibitem[{{Hjellming} \& {Han}(1995)}]{hjha1995}
{Hjellming}, R. \& {Han}, X. 1995, in X-ray binaries (Cambridge Astrophysics
  Series, Cambridge, MA: Cambridge University Press, |c1995, edited by Lewin,
  Walter H.G.; Van Paradijs, Jan; Van den Heuvel, Edward P.J.), p. 308

\bibitem[{{Hjellming} {et~al.}(1990{\natexlab{a}}){Hjellming}, {Han},
  {Cordova}, \& {Hasinger}}]{hjhaco1990}
{Hjellming}, R.~M., {Han}, X.~H., {Cordova}, F.~A., \& {Hasinger}, G.
  1990{\natexlab{a}}, \aap, 235, 147

\bibitem[{{Hjellming} {et~al.}(1990{\natexlab{b}}){Hjellming}, {Stewart},
  {White}, {Strom}, {Lewin}, {Hertz}, {Wood}, {Norris}, {Mitsuda}, {Penninx},
  \& {van Paradijs}}]{hjstwi1990}
{Hjellming}, R.~M., {Stewart}, R.~T., {White}, G.~L., {et~al.}
  1990{\natexlab{b}}, \apj, 365, 681

\bibitem[{{Homan} {et~al.}(1998){Homan}, {van der Klis}, {Wijnands}, {Vaughan},
  \& {Kuulkers}}]{hovawi1998}
{Homan}, J., {van der Klis}, M., {Wijnands}, R., {Vaughan}, B., \& {Kuulkers},
  E. 1998, \apjl, 499, L41

\bibitem[{{Hua} \& {Titarchuk}(1995)}]{huti1995}
{Hua}, X. \& {Titarchuk}, L. 1995, \apj, 449, 188+

\bibitem[{{Iaria} {et~al.}(2001){Iaria}, {Burderi}, {di Salvo}, {La Barbera},
  \& {Robba}}]{iabudi2001}
{Iaria}, R., {Burderi}, L., {di Salvo}, T., {La Barbera}, A., \& {Robba}, N.~R.
  2001, \apj, 547, 412

\bibitem[{{in 't Zand} {et~al.}(1999){in 't Zand}, {Verbunt}, {Strohmayer},
  {Bazzano}, {Cocchi}, {Heise}, {van Kerkwijk}, {Muller}, {Natalucci}, {Smith},
  \& {Ubertini}}]{invest1999}
{in 't Zand}, J.~J.~M., {Verbunt}, F., {Strohmayer}, T.~E., {et~al.} 1999,
  \aap, 345, 100

\bibitem[{{Jahoda} {et~al.}(1996){Jahoda}, {Swank}, {Giles}, {Stark},
  {Strohmayer}, {Zhang}, \& {Morgan}}]{jaswgi1996}
{Jahoda}, K., {Swank}, J.~H., {Giles}, A.~B., {et~al.} 1996, \procspie, 2808,
  59

\bibitem[{{Jonker} {et~al.}(2002){Jonker}, {van der Klis}, {Homan}, {M{\'
  e}ndez}, {Lewin}, {Wijnands}, \& {Zhang}}]{jovaho2002}
{Jonker}, P.~G., {van der Klis}, M., {Homan}, J., {et~al.} 2002, \mnras, 333,
  665

\bibitem[{{Jonker} {et~al.}(2000){Jonker}, {van der Klis}, {Wijnands}, {Homan},
  {van Paradijs}, {M{\'e}ndez}, {Ford}, {Kuulkers}, \& {Lamb}}]{jovawi2000}
{Jonker}, P.~G., {van der Klis}, M., {Wijnands}, R., {et~al.} 2000, \apj, 537,
  374

\bibitem[{{Kellermann} \& {Pauliny-Toth}(1969)}]{kepa1969}
{Kellermann}, K.~I. \& {Pauliny-Toth}, I.~I.~K. 1969, \apjl, 155, L71+

\bibitem[{{Klein-Wolt} {et~al.}(2002){Klein-Wolt}, {Fender}, {Pooley},
  {Belloni}, {Migliari}, {Morgan}, \& {van der Klis}}]{klfepo2002}
{Klein-Wolt}, M., {Fender}, R.~P., {Pooley}, G.~G., {et~al.} 2002, \mnras, 331,
  745

\bibitem[{{Matsuba} {et~al.}(1995){Matsuba}, {Dotani}, {Mitsuda}, {Asai},
  {Lewin}, {van Paradijs}, \& {van der Klis}}]{madomi1995}
{Matsuba}, E., {Dotani}, T., {Mitsuda}, K., {et~al.} 1995, \pasj, 47, 575

\bibitem[{{Migliari} {et~al.}(2003){Migliari}, {Fender}, {Rupen}, {Jonker},
  {Klein-Wolt}, {Hjellming}, \& {van der Klis}}]{miferu2003}
{Migliari}, S., {Fender}, R., {Rupen}, M., {et~al.} 2003, \mnras, in press,
  astro-ph/0305221

\bibitem[{{Mirabel} {et~al.}(1998){Mirabel}, {Dhawan}, {Chaty}, {Rodriguez},
  {Marti}, {Robinson}, {Swank}, \& {Geballe}}]{midhch1998}
{Mirabel}, I.~F., {Dhawan}, V., {Chaty}, S., {et~al.} 1998, \aap, 330, L9

\bibitem[{{Muno} {et~al.}(2001){Muno}, {Remillard}, \&
  {Chakrabarty}}]{murech2001}
{Muno}, M.~P., {Remillard}, R.~A., \& {Chakrabarty}, D. 2001, \apj, submitted,
  astro-ph/0111370

\bibitem[{{Penninx}(1989)}]{pe1989}
{Penninx}, W. 1989, in Hunt J., Battrick B., eds, 23rd ESLAB Symp. on Two
  Topics in X-Ray Astronomy, ESA SP-296, Volume 1: X Ray Binaries,, 185--196

\bibitem[{{Penninx} {et~al.}(1988){Penninx}, {Lewin}, {Zijlstra}, {Mitsuda}, \&
  {van Paradijs}}]{pelezi1988}
{Penninx}, W., {Lewin}, W.~H.~G., {Zijlstra}, A.~A., {Mitsuda}, K., \& {van
  Paradijs}, J. 1988, \nat, 336, 146

\bibitem[{{Penninx} {et~al.}(1993){Penninx}, {Zwarthoed}, {van Paradijs}, {van
  der Klis}, {Lewin}, \& {Dotani}}]{pezwva1993}
{Penninx}, W., {Zwarthoed}, G.~A.~A., {van Paradijs}, J., {et~al.} 1993, \aap,
  267, 92

\bibitem[{{Pooley} \& {Fender}(1997)}]{pofe1997}
{Pooley}, G.~G. \& {Fender}, R.~P. 1997, \mnras, 292, 925

\bibitem[{{Rothschild} {et~al.}(1998){Rothschild}, {Blanco}, {Gruber},
  {Heindl}, {MacDonald}, {Marsden}, {Pelling}, {Wayne}, \& {Hink}}]{roblgr1998}
{Rothschild}, R.~E., {Blanco}, P.~R., {Gruber}, D.~E., {et~al.} 1998, \apj,
  496, 538

\bibitem[{{Schnerr} {et~al.}(2003){Schnerr}, {Reerink}, {van der Klis},
  {Homan}, {M{\' e}ndez}, {Fender}, \& {Kuulkers}}]{screva2003}
{Schnerr}, R.~S., {Reerink}, T., {van der Klis}, M., {et~al.} 2003, \aap, 406,
  221

\bibitem[{{Schulz} {et~al.}(1989){Schulz}, {Hasinger}, \&
  {Truemper}}]{schatr1989}
{Schulz}, N.~S., {Hasinger}, G., \& {Truemper}, J. 1989, \aap, 225, 48

\bibitem[{{Sidoli} {et~al.}(2002){Sidoli}, {Parmar}, {Oosterbroek}, \&
  {Lumb}}]{sipaoo2002}
{Sidoli}, L., {Parmar}, A.~N., {Oosterbroek}, T., \& {Lumb}, D. 2002, \aap

\bibitem[{{Stella} {et~al.}(1985){Stella}, {White}, \& {Taylor}}]{stwhta1985}
{Stella}, L., {White}, N.~E., \& {Taylor}, B.~G. 1985, in Recent Results on
  Cataclysmic Variables, 125

\bibitem[{{Tan} {et~al.}(1992){Tan}, {Lewin}, {Hjellming}, {Penninx}, {van
  Paradijs}, {van der Klis}, \& {Mitsuda}}]{talehj1992}
{Tan}, J., {Lewin}, W.~H.~G., {Hjellming}, R.~M., {et~al.} 1992, \apj, 385, 314

\bibitem[{{Titarchuk}(1994)}]{ti1994}
{Titarchuk}, L. 1994, \apj, 434, 570

\bibitem[{{Ueda} {et~al.}(2001){Ueda}, {Asai}, {Yamaoka}, {Dotani}, \&
  {Inoue}}]{ueasya2001}
{Ueda}, Y., {Asai}, K., {Yamaoka}, K., {Dotani}, T., \& {Inoue}, H. 2001,
  \apjl, 556, L87

\bibitem[{{van der Klis}(1995{\natexlab{a}})}]{va1995a}
{van der Klis}, M. 1995{\natexlab{a}}, in X-ray binaries (Cambridge
  Astrophysics Series, Cambridge, MA: Cambridge University Press, |c1995,
  edited by Lewin, Walter H.G.; Van Paradijs, Jan; Van den Heuvel, Edward
  P.J.), p. 252

\bibitem[{{van der Klis}(1995{\natexlab{b}})}]{va1995b}
{van der Klis}, M. 1995{\natexlab{b}}, in Proceedings of the NATO Advanced
  Study Institute on the Lives of the Neutron Stars, held in Kemer, Turkey,
  August 29-September 12, 1993. Editors, M.A. Alpar, U. Kiziloglu, and J. van
  Paradijs; Publisher, Kluwer Academic, Dordrecht, The Netherlands, Boston,
  Massachusetts, p. 301

\bibitem[{{Wijnands} {et~al.}(1997){Wijnands}, {van der Klis}, {Kuulkers},
  {Asai}, \& {Hasinger}}]{wivaku1997}
{Wijnands}, R.~A.~D., {van der Klis}, M., {Kuulkers}, E., {Asai}, K., \&
  {Hasinger}, G. 1997, \aap, 323, 399

\bibitem[{{Zhang} {et~al.}(1993){Zhang}, {Giles}, {Jahoda}, {Soong}, {Swank},
  \& {Morgan}}]{zhgija1993}
{Zhang}, W., {Giles}, A.~B., {Jahoda}, K., {et~al.} 1993, \procspie, 2006, 324

\end{thebibliography}

\end{document}